# Nanostructuring strategies for silicon-based anodes in Lithium-ion batteries: Tuning areal silicon loading, SEI formation/irreversible capacity loss, rate capability retention and electrode durability

Dr. Mariam Ezzedine[a], Dr. Fatme Jardali[a], Dr. Ileana Florea[a], Dr. Mihai-Robert Zamfir[a,b], Prof. Dr. Costel-Sorin Cojocaru[a,*]

[a] Laboratory of Physics of Interfaces and Thin Films (LPICM), CNRS, École Polytechnique, IP Paris, 91128, Palaiseau Cedex, France.

[b] National Institute for Laser, Plasma & Radiation Physics (INFLPR), Atomistilor Street, No. 409, Magurele, Ilfov RO-077125, Romania.

*Corresponding author: costel-sorin.cojocaru@polytechnique.edu

## Batteries & Supercaps 2022

**Abstract**

Silicon is one of the most promising anode materials for Lithium-ion batteries. Silicon endures volume changes upon cycling, which leads to subsequent pulverization and capacity fading. These drawbacks lead to a poor lifespan and hamper the commercialization of silicon anodes. In this work, a hybrid nanostructured anode based on silicon nanoparticles (SiNPs) anchored on vertically aligned carbon nanotubes (VACNTs) with defined spacing to accommodate volumetric changes is synthesized on commercial macroscopic current collector. Achieving electrodes with good stability and excellent electrochemical properties remain a challenge. Therefore, we herein tune the active silicon areal loading either through the modulation of the SiNPs volume by changing the silicon deposition time at a fixed VACNTs carpet length or through the variation of the VACNT length at a fixed SiNPs volume. The low areal loading of SiNPs improves capacity stability during cycling but triggers large irreversible capacity losses due to the formation of the solid electrolyte interphase (SEI) layer. By contrast, higher areal loading electrode reduces the quantity of the SEI formed, but negatively impacts the capacity stability of the electrode during the subsequent cycles. A higher gravimetric capacity and higher areal loading mass of silicon is achieved via an increase of VACNTs carpet length without compromising cycling stability. This hybrid nanostructured electrode shows an excellent stability with reversible capacity of 1330 mAh g$^{-1}$ after 2000 cycles.

**Introduction**

Lithium-ion batteries (LIBs) based on layered metal oxide cathodes and graphite anodes have supplanted other rechargeable batteries and are nowadays considered the most auspicious energy storage systems due to their high energy and power densities.[1] LIBs have been extensively used in portable devices and electric vehicles and are potential candidates for sedentary energy storage applications.[2,3] However, commercial LIBs are now approaching their specific energy density limits.[4] The currently achieved cell-level gravimetric energy density is approximately 250 Wh kg$^{-1}$ at a price of ~ 150 USD kWh$^{-1}$.[5] The continuous increase in energy consumption and the perpetual higher performance demand in various application fields requires technological breakthroughs in the current battery technology towards achieving higher energy density LIBs (>400 Wh/kg).[6]

The exploration of novel battery configurations for enhancing the energy density of the cell is generally practiced through electrode architecture engineering that targets increased active material loading per unit area. Higher active material loading can be achieved by increasing electrode thickness and/or boosting the active material fraction. [7] However, increasing the electrode thickness in conventional electrode configurations, i.e., where the anode/cathode is slurry coated on metallic current collectors, remains challenging. Increasing the thickness of electrodes proportionally increases the charge transport distance, i.e., electrons and Li$^+$ ions require more time to diffuse throughout the entire electrode, resulting in increased cell resistance, degraded rate capacity and a limited enhancement in energy density.[8] The randomness and close packing of the components in a conventional slurry coated electrode generally leads to a torturous pore structure, impeding optimum infiltration of the electrolyte and considerably increasing the ion transfer distance. According to the effective ionic conductivity expression, defined as $D_{int} = \frac{\epsilon}{\tau} D_{int}$ (where $\epsilon$ is the porosity, $\tau$ is the tortuosity, and $D_{int}$ is the intrinsic ion conductivity), $D_{eff}$ is inversely proportional to the tortuosity of the electrode.[9] Therefore, designing electrodes with low-tortuosity pore structures is fundamental for thick electrode architectures in order to improve the ion transfer rate. Moreover, the fabrication of robust thick electrode layers remains a great challenge as thick electrodes tend to fracture and/or delaminate from the current collector during the drying process.[10] Additionally, one must consider that the power density is often equally important for thick electrode designs. High power electrodes demand a high porosity for ion transport and a large volume of conductive fillers for insuring efficient electron transport.[8]

Many studies have been focused on accelerating charge transport by optimizing the ion- and electron-conductive network inside the electrodes through the development of novel conductive fillers and by establishing an interconnected conducting network within the electrode.[11] Complementing the conventional carbon materials, such as carbon black and graphene powder, that are commonly used in commercial batteries,[10] carbon fibers and graphene nanosheets have emerged as attractive conductive additives due to their high electrical conductivities and their effectiveness in constructing nanoscale circuitry throughout the electrode structure.[12] Unfortunately, issues associated with the low tap density, low Coulombic efficiency, and prolonged charge-transfer distance have also accompanied the introduction of such advanced carbon fillers into thick electrodes, thereby hindering the ability to achieve high energy and power densities in LIBs.[13] Recently, self- or hierarchically- assembly approaches have been proposed to accomplish synergies between primary carbon nanostructures and advanced architectures in high performance electrodes.[14]

Carbon nanotubes (CNTs) vertically aligned on the surface of a current collector, also called "CNT carpet" or VACNTs, match ideally with the specifications needed in the field of LIBs. The vertical alignment of CNTs in the direction of the ion diffusion provides an increase in both the storage capacity (via increased effective achievable electrode thickness) and the rate capability. VACNTs ensure better contact and adhesion with the substrate, resulting in an increase in the ion diffusivity as well as an improvement of the electron transport properties, which allow improvements in the rate capabilities.[15] Furthermore, the intimate contact between each of the VACNTs and the substrate provides a low resistance pathway allowing for an effective connectivity throughout the electrode. Finally, VACNTs exhibit a porous structure with a large surface area, which greatly facilitate the chemical transport and the interfacial reaction. [16–18]

VACNTs possess other merits, including i) enhanced specific surface due to the high CNTs density of $10^{11}$ – $10^{13}$ cm$^{-2}$ [19] combined with macro heights (up to centimeters), ii) high degree of alignment of individual tubes achieved via a steric effect (i.e., via van der Waals forces between neighboring CNTs forcing all tubes to grow vertically) without requiring any post-processing steps, and iii) the control on the intrinsic properties of the CNTs (wall number, tube diameter, and density) via the initial catalyst layer thickness , its chemical nature (Fe, Co, Ni or alloys) and the duration of the pre-treatment step.[20] The VACNTs thus emerge as extremely attractive potential candidates for electrode materials or as current collectors in high performance LIBs.

Owing to its abundancy and its high theoretical capacity of 4200 mAh g$^{-1}$ when forming $Li_{22}Si_5$ and 3580 mAh g$^{-1}$ when forming $Li_{15}Si_4$,[21] silicon (Si) is regarded as the next generation anode material for LIBs. Nevertheless, the fast capacity fading and reduced Coulombic efficiency resulting from the huge volume expansion and shrinkage due to its alloying mechanism with Li (up to 320%) and the cumulative and unstable solid electrolyte interphase (SEI) formation hinder the silicon-based anode material for further practical applications.[22] Anodes in the form of low-dimensional silicon structures (0D or 1D) have several advantageous characteristics compared with bulky silicon anodes including improved electrochemical behaviors due to the smaller silicon size resulting in better accommodation with volume changes as well as better rate performance due to larger surface area resulting in fast electron transfer and shortened lithium diffusion pathway.[23] Among the different systems, nanowires (NWs) provide a highly porous medium, which allows for an easy Si expansion upon alloying with Li. Moreover, SiNWs can be directly connected to the anode current collector, i.e., neither a binder nor conductive additives are needed for fabricating the anode.[24] However, increasing the areal mass loading (mg cm$^{-2}$) of vapor-liquid-solid-grown SiNWs-based electrodes remains a challenge. Typical mass loadings of several mg cm$^{-2}$ of active material, i.e., ~10 times more than those presently obtained should be reached.[25] Amorphization is also reported in SiNW-based anodes as the electrochemical insertion of Li$^+$ ions at room temperature directly destroys the Si crystal structure leading to the formation of a metastable amorphous Li–Si alloy.[26]

Much attention has been directed towards the exploitation of the combined merits of two nanoscale materials: Si nanoparticles (SiNPs) and CNTs. Wang *et al* designed a structure composed of carbon-coated SiNPs anchored on CNTs and confined in cellulose carbon rolls. The designed microscroll structure represent a crossover from the slurry structure used in commercial electrodes to a 1D electrode architecture. This structure showed high electrode specific capacity and good stability (2000 mAh g$^{-1}$ after 300 cycles).[27] Gohier *et al.* previously reported a hierarchical hybrid nanostructure based on VACNTs directly grown onto stainless steel foil on which SiNPs were deposited on the VACNTs' walls (SiNPS@VACNTs) *via* a Chemical Vapor Deposition (CVD) process.[14] In their work, the deposition duration of SiNPs was 13 minutes on short VACNT carpet (~10 μm) that allowed to obtain areal loading of 0.17 mg cm$^{-2}$. Their results showed that such electrode configuration exhibits an impressive electrochemical performance of 760 mAh g$^{-1}$ at 15 C, highlighting the potential of such a hierarchical nano-architecture anode. The key factors for the excellent cycling properties are the perfect adhesion between CNTs (direct connection to the macroscopic current collector) and silicon particles as well as the low tortuosity of the electrode structure facilitating electron and lithium ion transport pathway and limiting the diffusion process occurring in conventional electrodes.

The present work provides a strategy to modulate the performance of a hierarchically assembled SiNPs@VACNTs as an active anode electrode, and attain a high electrode areal capacity, high cycling stability, along with excellent electrical conductivity and adjustable electrode thickness. Using Si in its nanostructured form grafted on VACNTs nanostructured current collector will allow to accommodate huge volumetric changes upon cycling while maintaining the structural integrity of the electrode. With this objective, systematic experiments are realized to tune the areal SiNPs loading in the electrode following two approaches. The first approach consists of controlling the SiNPs volume or size by controlling the Si deposition time at a fixed VACNTs carpet length. The second approach involves the adjustment of the VACNTs carpet length at a fixed SiNPs volume. Moreover, the present study provides an approach to modulate SEI formation and irreversible capacity losses. This strategy offers a way to design 1D electrodes with good stability and excellent electrochemical properties for high performance batteries.

**Results and discussion**

**Morphological and structural characterizations**

**VACNT carpets with different lengths**

Scanning Electron microscopy images of VACNTs grown on Cu foils with different carpet lengths, i.e., between 27 and 145 µm are shown on Figure 1. The dHF-CVD approach also enables the simultaneous growth of VACNTs on both sides of the Cu foil as shown in Figure S2. The growth parameters chosen in this study yield uniform VACNT carpets comprised of CNT with 3 to 6 walls and outer diameter of 5 to 8 nm as shown in Figure 1e, and with CNTs densities in the range of $10^{11}$ -$10^{12}$ tubes cm$^{-2}$. In general, the diameter, number of walls and density of CNTs can be tuned via the initial catalyst layer thickness, its chemical nature (such as Fe, Co, Ni or alloys) and the duration of the pre-treatment step with activated hydrogen.

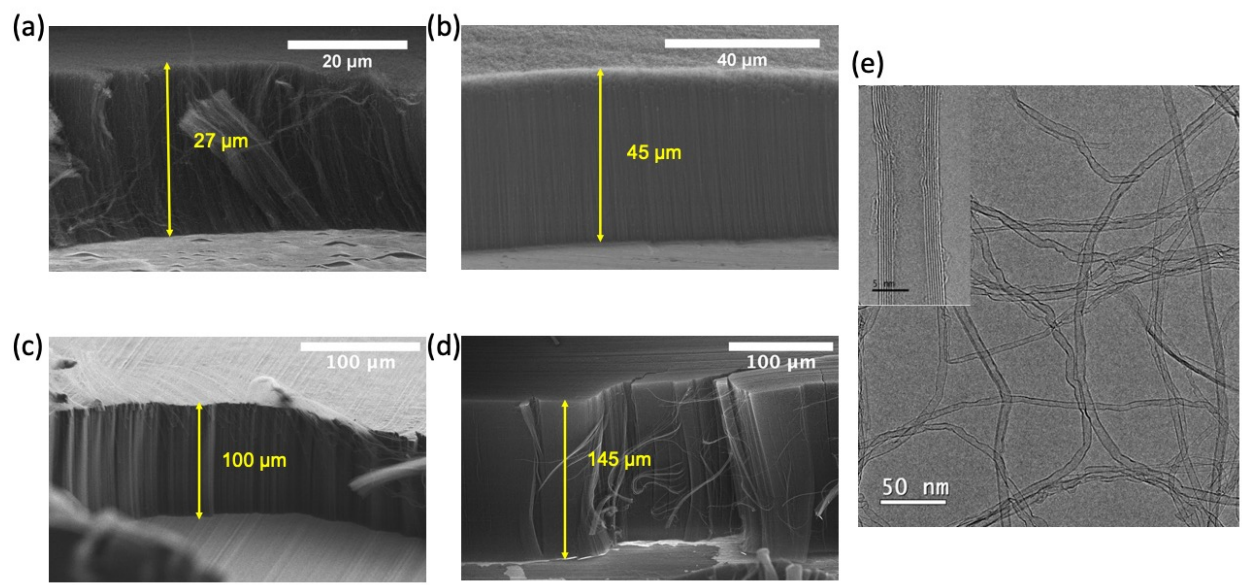

Figure 1. Cross-sectional SEM images of VACNTs synthesized on Cu foil. The length of the VACNTs carpet is 27 µm (a), 45 µm (b), 100 µm (c) and 145 µm (d). TEM observation of the CNTs (e); HRTEM image illustrating the tube diameter size and number of walls (inset).

**Hybrid SiNPs@VACNTs nanostructured electrodes**

To increase the areal mass of Si deposited on the electrode, two approaches were explored. The first one consists in increasing the deposition time for SiNPs formation while keeping the height of the VACNTs carpet fixed at 30 μm. The second approach consists in fixing the deposition time of SiNPs on the nanotube sidewalls for 5 minutes and modulating the total areal number of SiNPs by growing various lengths of VACNTs carpets. Accounting for the porosity of the CNT carpets, the main advantage of this second approach is that it allows one to achieve a higher overall electrode areal mass of Si while avoiding the formation of a thin silicon layer encompassing the tubes and, thus, prevents the undesired cracking of the electrode.

Cross-sectional SEM micrographs of the resulting SiNPs@VACNTs heterostructure assemblies obtained on 30 μm VACNT carpets and at two different Si deposition durations, i.e., 5 and 10 minutes, are presented in Figure 2a and 2c, respectively. Figures 2b and 2d show the high magnification cross-sectional SEM micrographs of the hybrid SiNPs@VACNTs nanostructure assemblies obtained for 5 and 10 minutes, respectively, which distinctly show the SiNPs attached to the individual CNTs. The overall integrity of the CNT carpets remains undamaged after Si deposition. As depicted in Figure 2b and 2d, the CNT surface supplies the desired high-density sites for SiNPs nucleation, growth, and homogeneous coverage of the VACNTs during the subsequent $SiH_4$ decomposition step. It is noticeable for the 5 minutes deposition the presence of a significantly large inter-particle spacing which can accommodate the expected stress resulting from the volume expansion during the lithiation phase. As expected, with an increase in the deposition time, the morphology of the nanoparticles evolves from small SiNPs coverage towards bigger SiNPs covering entirely the CNTs surface. Figures 2e and 2g present SEM micrographs of the hybrid SiNPs@VACNTs nanostructured assemblies for two different lengths of the VACNTs carpets, i.e., 30 and 60 μm, respectively, which distinctly show the SiNPs attached to the CNTs. In this case, the overall configuration of the CNT carpets also remains intact after Si deposition. Regardless their length, the entire CNT surface provides the desired sites for the SiNPs to nucleate and grow during the subsequent $SiH_4$ decomposition step, thus, allowing for a highly homogeneous CNTs coverage with SiNPs.

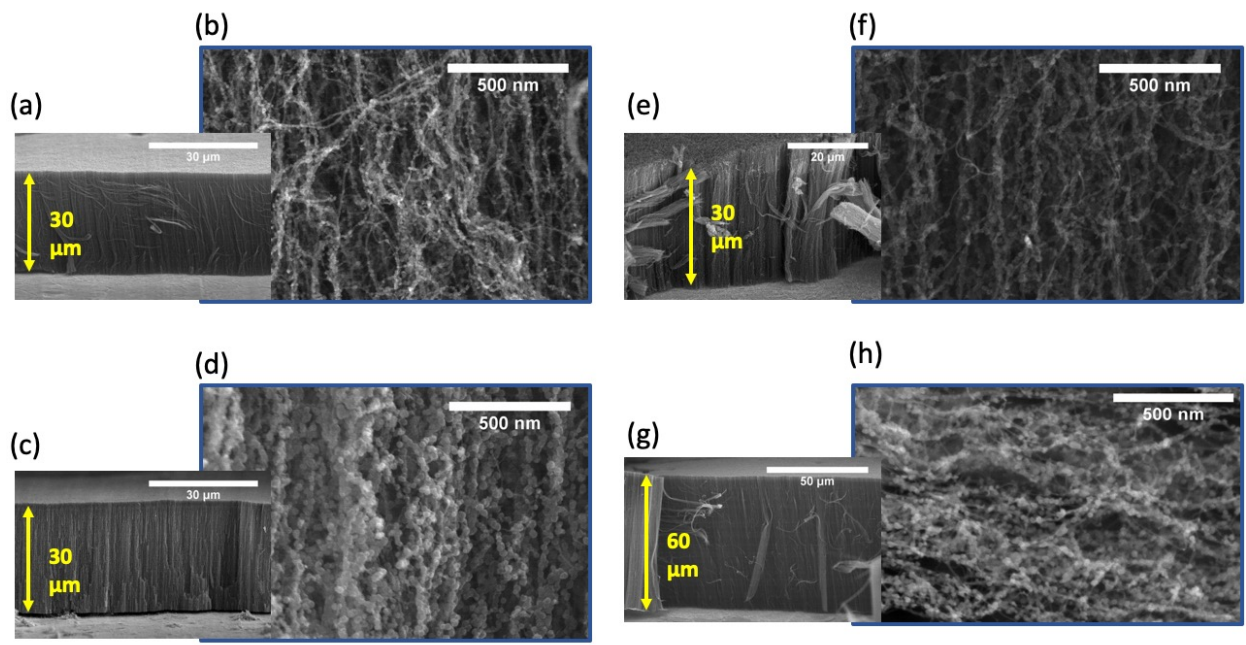

Figure 2. SEM analyses of SiNPs@VACNTs nanostructure assemblies for the two proposed approaches: SEM micrograph and its corresponding high magnification obtained after 5 minutes (a) and (b) and 10 minutes (c) and (d) exposure to a SiH₄ environment. SEM micrographs and its corresponding high magnification obtained for different lengths of the VACNTs carpet; 30 μm (e) and (f) and 60 μm (g) and (h), respectively. In the inset, a high magnification SEM micrograph illustrating a homogenous Si deposition.

The SEM macroscopic observations were furthermore confirmed by the TEM analysis performed in both TEM and STEM imaging modes. As can be seen in Figures 3a and 3b after 5 minutes under a SiH₄ environment, amorphous SiNPs form on the nanotube sidewalls exhibiting a rather spherical morphology and a homogenous size. The average SiNPs size, determined by the statistical analysis of more than 350 NPs, is 15 ±2.5 nm. The SiNPs size increases to 40 nm when the deposition time is increased to 10 minutes, as shown in Figure 3c. Regarding the SiNPs localization on the CNT sidewalls, our previous studies using electron tomography technique helped us assessing their 3D localization which can reasonably be correlated to the presence of point-defects on the external graphitic structure of the tube.[28] These defects, formed during the nanotube growth, can be considered as anchorage points for the Si precursors that nucleate into seeds and start increasing their volume to form SiNPs by continuous local SiH₄ gas decomposition. This morphology evolution also induces a modification of the overall SiNPs specific surface area, which tends to decrease with the increase of the deposition time, as it is clearly evidenced by comparing the 5 minutes deposited samples with the 10 minutes deposited ones.

High resolution TEM analyses of SiNPs evidence their amorphous structure as evidenced by the FFT and X-ray diffraction spectrum (Figure S2). Further details of their composition were obtained by the HAADF. Figure 3d shows a STEM-HAADF image of a nanostructured SiNPs@VACNTs assembly together with its corresponding Si/O relative map obtained by superimposing the two elemental maps (see inset) which provides a complete image on the distribution of the two principal residual components, Si and O, within the NPs. Elemental maps were extracted at energies of 1.74 keV (Si Kα) and 0.53 keV (O Kα). As one can observe, the NPs content is mainly of elemental Si enclosed by a thin O layer (2 nm) at the surface which most probably appeared due to ambient exposure during the transfer from the CVD reactor to the microscope (see Figure 3(e-g)). The closer analysis done by STEM-HAADF EDS line scan across a region containing two silicon nanoparticles attached to the walls of the CNTs (see Figure S3) indicates the presence of carbon coming from the tubes or from the presence of some amorphous carbon. Since there is no overlap between the carbon and silicon line profiles, the formation of silicon carbide can be excluded.

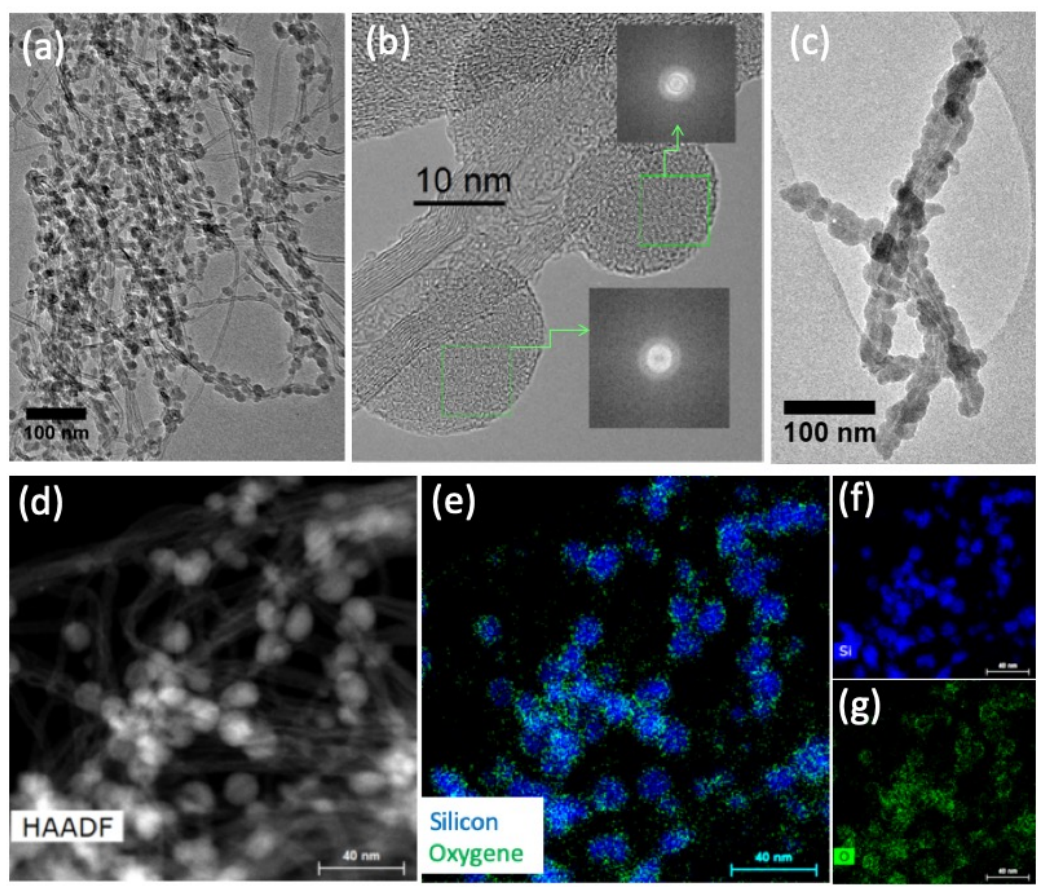

Figure 3. TEM and STEM observations of SiNPs@VACNTs nanostructures synthetized for different exposure times under SiH$_4$ environment: 5 minutes (a) and 10 minutes (c). (b) High resolution TEM image on two SiNPs from (a) and their corresponding FFT, evidencing their amorphous structure. (d-g) STEM-HAADF-EDX elemental map on a large area of nanostructured SiNPs@CNTs assemblies obtained for 5 minutes exposure to SiH$_4$ gas: (d) STEM-HAADF image of SiNPs@CNTs assembly; (e) corresponding Si/O relative map with Si represented in dark blue (f) and O in green (g).

**Electrochemical characterizations of SiNPs@VACNTs electrodes at various Si loading mass**

The gravimetric capacities of the SiNPs@VACNTs electrodes were calculated by considering solely the effective mass of deposited Si. This value was determined by measuring the weight before and after Si deposition on the VACNT carpets for the two approaches. The VACNTs were not accounted for calculating the theoretical capacity of the electrode. Thus, for the samples considered in the first approach (variable SiNPs deposition time), the Si loading mass obtained for 5- and 10-minutes deposition were 0.25 and 2.5 mg cm$^{-2}$, respectively. For the second approach (variable VACNT carpet length), the mass of Si increases from 0.2 to 0.8 mg cm$^{-2}$ with the increase of the carpet length from 30 to 60 μm at Si deposition time of 5 minutes. The samples were mounted in coin cells inside a glove box and were tested vs Li foils in a half-cell configuration in a voltage window of 2 V to 20 mV at a C/20 rate for the first few cycles in order to stabilize the structure and then the cycling was pursued at a 1C for both approaches.

Before testing lithiation and delithiation behavior of Si@VACNTs as active electrode, we evaluated the electrochemical contribution of pristine VACNTs carpet and compared it to Si@VACNTs electrode. We applied similar C rate conditions for both samples (1$^{st}$ cycle at C/20 and subsequent cycles at 1C). Figure S4 in Supporting Information show the potential profiles of 1$^{st}$ (in inset), 2$^{nd}$, 5$^{th}$ and 10$^{th}$ cycles versus capacity (mAh) of pristine VACNTs (a) and Si@VACNTs (b). During the first cycle, the capacities for pristine VACNTs reach 1.5 mAh and 0.1 mAh for lithiation and delithiation, respectively. The shape of the reduction curve is typical of SEI formation onto carbon nanotubes, in good agreement with literature data on CNTs electrochemistry.[29] We notice rapid and continuous drop in the lithiation capacity during the subsequent cycles where the capacities remain in the μAh range. When Si is deposited on VACNTs, a high capacity of 1.9 mAh is obtained for the 1$^{st}$ lithiation cycle where the shape of the curve indicates that the formation of SEI between 2V and 0.2V vs Li$^+$/Li as well as the plateau of lithium alloying process at about 0.2V. The capacities of Si@VACNTs obtained during the subsequent cycles remain stable at about 0.9 mAh, which is forty-fold higher than pristine VACNTs.

**Tuning areal silicon loading, SEI formation and irreversible capacity loss**

**Modulating SEI formation and irreversible loss in accordance with SiNPs volume**

Before discussing the detailed electrochemical performance of the SiNPs@VACNTs electrodes, we will focus on some insights concerning the specific differences of the SEI formation on such hierarchically nanostructured anodes as compared to commercial graphite anodes. It is well known that lithiation and delithiation of Si take place at low voltage, inducing the formation of the SEI film on the anode surface due to the decomposition of the electrolyte solvents.[30,31] The specific area of the active material in contact with the electrolyte is, therefore, expected to play an important role on both the formation and the stability of the SEI layer. Increasing SiNPs deposition time translate into formation of larger size nanoparticles thus a decrease of the specific area of the active material in contact with the electrolyte is expected. We have investigated the SEI formation on such nanostructured anodes for 5- and 10-minutes deposition times of SiNPs. From the galvanostatic curve of the sample with the highest specific surface area prepared at 5 minutes of SiNPs deposition shown in the Figure 4a, a gentle sloping plateau starting at 1.3 V is observed, and we ascribe it to the reduction of the fluoroethylene carbonate (FEC) additive from the electrolyte on the surface of the SiNPs.[32] The slight slope observed around 1.3 to 1.0 V can be attributed to the decomposition of the other components in the electrolyte solution, leading to the formation of the SEI layer. The Coulombic efficiency in the first cycle is 30%. The observed large irreversible capacity can be related to the SEI formation over the large specific area of the SiNPs@VACNTs in this nanostructured electrode, due to trapping of Li$^+$ ions into the SEI during its intensive formation. When the volume of the deposited SiNP is increased, it is expected that their corresponding surface area to decrease. Consequently, for the 10 minutes deposited sample shown in Figure 4b, the magnitude of the plateau corresponding to the SEI layer formation is clearly reduced, suggesting a diminished amount of SEI being formed on this electrode in accordance to a lower specific surface area of the structure. As a result, the Coulombic efficiency increases up to 67%. The improvement of the Coulombic efficiency and the diminishment of the plateau related to the decomposition of the electrolyte can both be reasonably explained by a decrease in the specific surface area of the active SiNPs decorating the VACNTs. The formation of the SEI layer is obviously still present, but it exhibits considerably alleviated irreversible capacity loss during the first cycle. The increase in the deposition time and, in turn, the increase in the NPs size up to bigger particles wrapping the VACNTs reduces the subsequent formation amount of the insulator SEI layer due to the decreased specific Si surface area, which in turn translate into a higher Coulombic efficiency. This is consistent with previous reports of the irreversible capacity loss of the SiNPs being linked to the side reactions between the electrolyte products and the Si material, leading to a continuous formation of the SEI layer and the trapping of Li$^+$ ions that are irreversibly consumed into the electrolyte decomposition.[33] These phenomena largely contribute to the observed severe capacity drop of the cell. A low irreversible capacity loss in the incipient cycles is of great importance when considering the potential integration of Si based anodes in a full LIB, as it will impact the ability to achieve proper balancing of the specific surface capacities of the anode and the cathode. Thus, these results provide us with important information to be considered when choosing the morphology of the nanostructured anode with respect to the desired areal loading mass. Comparing our results to literature, the first-cycle Coulombic efficiency of SiNPs with large specific surface is in the range of about 20-40% (Figure 4a), far below that of commercial graphite anodes (90-94%).[30] However, to overcome these issues, an increase in the Si content in the electrode can improve the first-cycle Coulombic efficiency. Many research efforts have been done to increase the initial Coulombic efficiency to above 80% by incorporating heteroatoms in carbon modifier and artificial graphite such as natural-biogum-deprived carbon composites Si@GAC, Si@GGC and Si@XGC anode materials, [34] adding nano-silver to spherical Si@SiOx nanocomposites anode, [35,36] and tuning core-shell Si@SiOx nanostructures.[37] Aside from the SEI layer formation, the electrochemical lithium storage characteristics of the SiNPs@VACNTs electrode are consistent with typical amorphous silicon lithiation/ delithiation behavior for the two approaches discussed below.

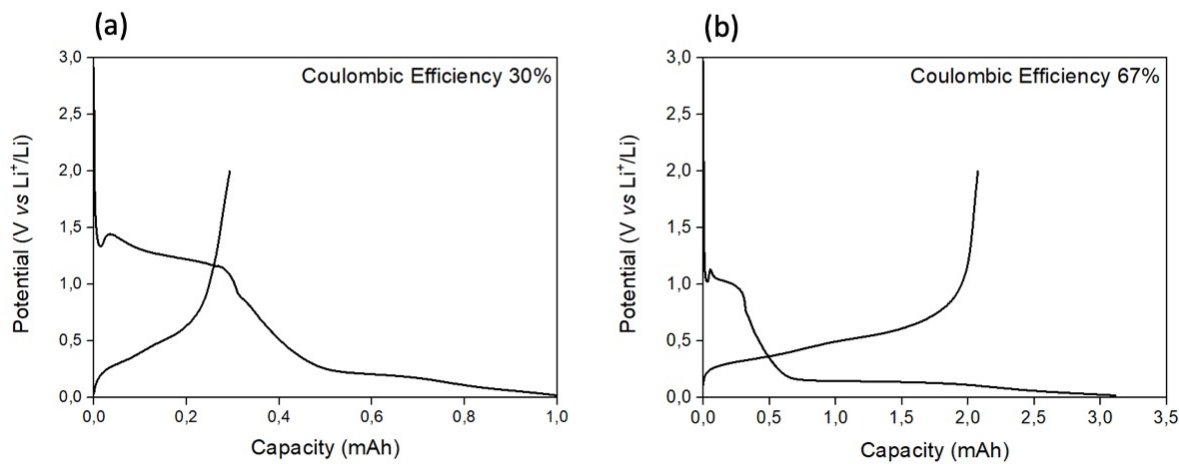

Figure 4. Galvanostatic curves of the first cycle for samples with different Si deposition times: (a) Lighter loaded electrode (5 min) and (b) higher loaded electrode (10 min).

**First approach for tuning Si areal loading: Controlling SiNPs volume via deposition time on 30 μm VACNTs carpet**

The lithiation / delithiation capacity plots for 40 cycles of the lighter and higher loaded SiNPs@VACNTs nanostructured anode of the first approach are shown in Figures 5a and 5b, respectively. For the lighter loaded electrode (i.e., 5 minutes of Si deposition), the second lithiation / delithiation areal capacities are 1.1 and 0.8 mAh cm$^{-2}$, respectively with a Coulombic efficiency of 72%, corresponding to an irreversible capacity loss of 28%. In the subsequent cycles, the Coulombic efficiency reaches 98-99% and remains relatively stable. The areal capacity of the SiNPs@VACNTs anode retains at 0.25 mAh cm$^{-2}$ after 40 cycles at 1C rate indicating the promising nature of this nanostructured anode which shows a higher capacity compared to the theoretical capacity of graphite. The cycling plot for 40 cycles for the higher loaded electrode (i.e., 10 minutes of Si deposition) is shown in Figure 5b. The second lithiation / delithiation capacities are 5.2 and 5.0 mAh cm$^{-2}$, respectively with a Coulombic efficiency of 91%, corresponding to an irreversible capacity loss of 9%. At the end of the 40 cycles, a capacity retention close to 1.2 mAh cm$^{-2}$ is obtained when cycled at a 1C rate. For the higher loaded sample, this specific capacity at 1C also suffers from a gradual fade of about 40% at the end of the 40$^{th}$ cycle, whilst the lighter loaded sample exhibit a stable specific capacity over the same number of cycles. The capacity fading can be explained by the excessive amount of large SiNPs present on the VACNT walls which probably had undergone cracking and pulverization of the electrode. These results suggest that although increased volume of the deposited SiNPs is beneficial both in terms of increased areal loading (for the same length of VACNT carpet) and in terms of reduced irreversible losses during initial SEI formation, there is an associated drawback of reduced capacity retention and increased capacity fading. Larger size SiNPs minimize the quantity of SEI being formed during the first cycles, however the associated increase of accumulated mechanical stress correlated with the increase of Li$^+$ insertion path (larger particles diameter) impact negatively both the electrode stability and its ability to accommodate higher cycling rates. These observations are consistent with previous reports in literature on other Si nanostructures.[24,38–40]

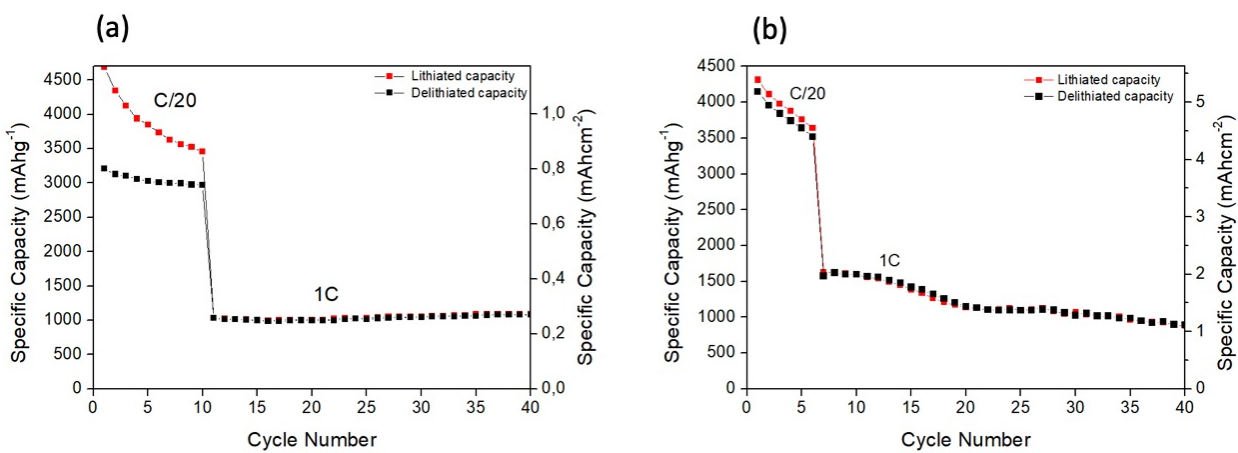

Figure 5. Lithiation / delithiation capacities of SiNPs@VACNTs electrodes during 40 cycles: (a) Lighter loaded electrode (5 min), (b) higher loaded electrode (10 min).

In Figure 6, we compare different loadings, i.e., Si deposition times of 5, 10 and 12 minutes on 30 μm VACNTs and we show the areal capacities for the first 6 cycles at C/20 rate and the SEM micrographs of the sample obtained after 12 minutes Si deposition (for comparison with different Si deposition time, refer to Figure 2b, 2d). When the deposition time is increased to 12 minutes, the Si loading mass is 4 mg cm$^{-2}$ with an areal capacity that reaches 7.2 mAh cm$^{-2}$ at the second cycle. We notice a capacity fading during the first cycles which can be explained in regard to the formation of very large particles of Si on the walls of the CNTs as clearly seen in the SEM image which presumably have undergone increased pulverization upon lithiation and delithiation of the electrode.

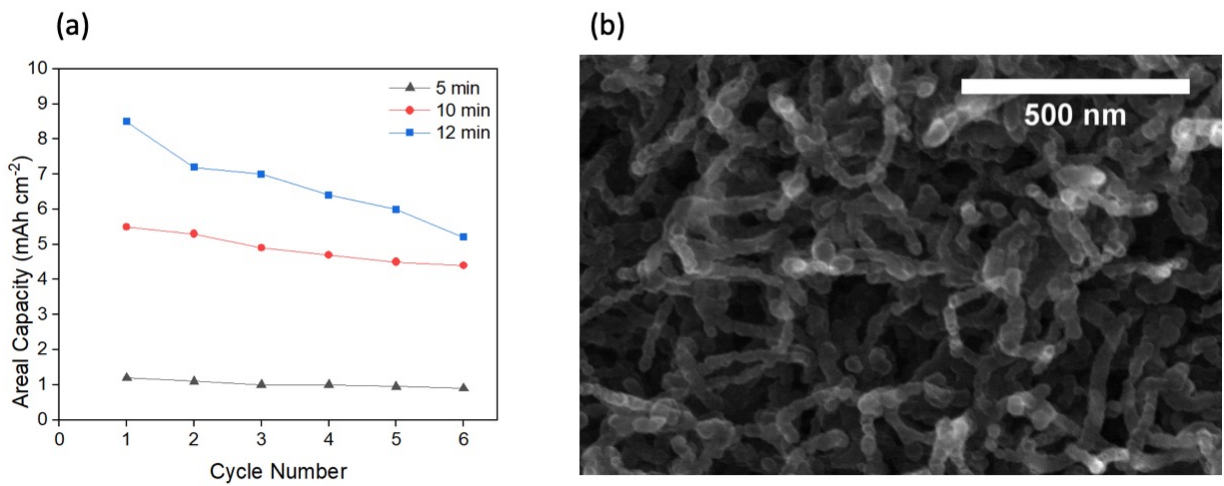

Figure 6. A comparison of the areal capacities during the first 6 cycles for 5, 10, and 12 minutes of Si deposition on 30 μm VACNTs (a). SEM analyses of SiNPs@VACNTs nanostructure obtained after 12 minutes deposition (b).

**Second approach for tuning Si areal loading: Adjusting VACNTs carpet length at a fixed SiNPs volume**

Proper tuning of the Si deposition time and consequently of the size of the decorated SiNPs appears as a convenient way for achieving desired balance between the increased Si areal loading with minimum irreversible losses during initial SEI formation cycle and the rate capacity of the electrode. However, for many practical applications higher rate capacity is simultaneously desired with a higher electrode areal loading.[10,41] In order to achieve such balance, for a fixed optimum SiNPs size, we propose to use increased length of the VACNTs. Assuming homogeneous deposition of the SiNPs on the CNTs surface, it should become possible in this case to increase Si areal loading without the negative impact neither on the capacity retention of the electrode, nor on its rate capacity. The electrochemical characteristics of the two samples with different areal loadings obtained by increasing the length of nanotubes were compared in Figures 7a and 7b. For the sample with the shorter VACNTs (i.e., 30 μm), the second lithiation / delithiation areal capacities were 0.8 mAh cm$^{-2}$ and 0.7 mAh cm$^{-2}$, respectively, for a Coulombic efficiency of 91% at C/20 rate. These values increased as the carpet length increased to 60 μm owing to the contribution from the higher content of SiNPs achieving a higher lithiation / delithiation areal capacities of 2.0 and 1.7 mAh cm$^{-2}$, respectively, for a Coulombic efficiency of 88%.

Comparing these two approaches, the results point out that overall the SEI layer formation behave similarly independent of the electrodes Si areal loading, as the SiNPs morphology is similar, resulting in similar SiNPs individual specific area. The electrochemical characteristics of the first sample are quite similar to those obtained in the previous study for the same VACNT length and same Si deposition time (i.e., 30 μm and 5 min) which is a positive indication of the robustness of our fabrication process. For this sample with a Si loading mass of 0.2 mg cm$^{-2}$, an areal capacity of about 0.2 mAh cm$^{-2}$ is obtained at 1C rate that is stable during the subsequent 40 cycles, similar to the results of the first sample cycled at the same C rate. In contrast, for the second sample (i.e., 60 μm) with a higher loading of Si, a reversible capacity of about 0.6 mAh cm$^{-2}$ was retained for 40 cycles at 1C rate. An approximately 9.5% loss of the specific capacity can be observed for the first sample (i.e., 30 μm) during the cycling at 1C. Remarkably, the capacity fade for the higher Si content electrode (i.e., 60 μm) represents about 21% loss at 1C. This behavior is in stark contrast with the higher Si loaded electrode in the first approach (i.e., large SiNPs on 30 μm VACNTs length) and provides strong indication of a limited accumulated mechanical stress upon lithiation and limited mechanical disintegration of the electrode due to the volume expansion. Importantly, these experimental results indicate that the height of the VACNTs carpet does not seem to limit neither the electronic nor the Li$^+$ ions transport in such nanostructured electrodes. In fact, higher VACNT length electrodes exhibited a higher gravimetric capacity than shorter ones at a given current as expected from the higher areal loading mass of Si.

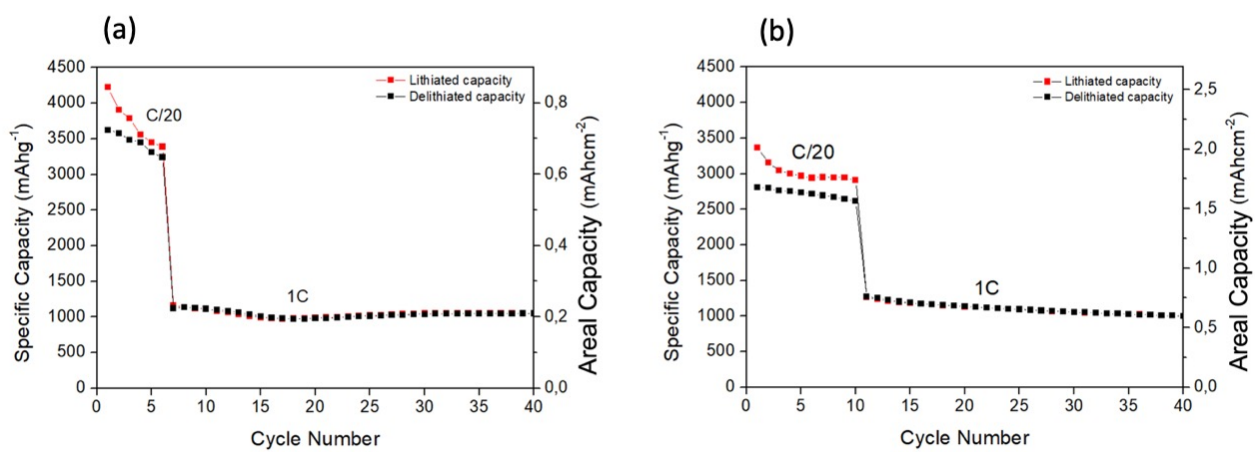

Figure 7. Lithiation / delithiation capacities of SiNPs@VACNTs electrodes during 40 cycles: (a) 30 μm VACNTs and (b) 60 μm VACNTs.

Table S1 summarizes the areal and specific capacities obtained in this work to previously reported silicon-based anodes. Figure 8 shows the effect of VACNTs length and the size of deposited SiNPs on the areal capacity based on theoretical estimations (refer to Supporting Information for more details). It is clear that an areal capacity greater than 8 mAh cm$^{-2}$ can be reached when the length of VACNTs carpet is increased to more than 100 μm while keeping the size of deposited Si to 20 nm which is in the range of the SiNPs size obtained after 5 minutes of deposition.

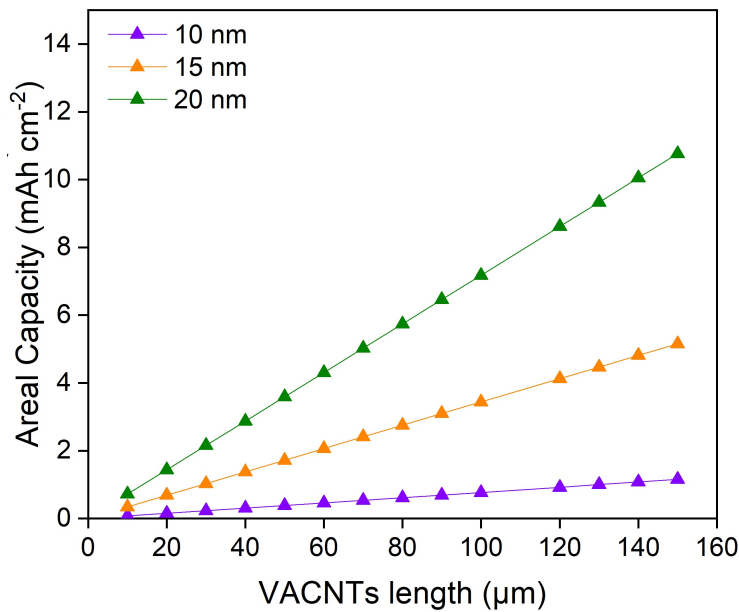

Figure 8. Estimation of the theoretical areal capacity obtained for different VACNT lengths and Si diameters.

**Electrode durability and rate capability retention of a lighter loaded SiNPs@VACNTs electrode**

**Electrode durability**

To highlight the excellent capacity retention of such nanostructured electrode upon long term cycling, and to provide a comparison of the performance with the state-of-the-art graphite anodes, we conducted thorough electrochemical testing of a lighter loaded SiNPs@VACNTs electrode with VACNT length of 30 μm (see Figure 9a). Testing began at a C/20 rate for the first 6 cycles with the purpose of stabilizing the structure and the SEI layer. Subsequently, the half-cell was cycled at relatively high rates and the capacity retention for over 2000 cycles was recorded. The second lithiation and delithiation capacities were 4091 and 2281 mAh g$^{-1}$, respectively, resulting in an initial Coulombic efficiency of 66%. During the subsequent cycles, the Coulombic efficiency increased gradually to reach 99% and was preserved as the cycles proceeded. After changing the C rate, we observed a slight increasing trend in the lithiation and delithiation capacities starting from the 19$^{th}$ cycle, which can be attributed to the activation of more Si atoms reacting with Li$^+$ ions.[42] Reversible capacities of 2380, 1731, and 1440 mAh g$^{-1}$ were retained for 100, 500 and 1000 cycles, respectively at C/5. After 2000 cycles, the hybrid hetero-structured SiNPs@VACNTs electrode retained a reversible capacity of about 1330 mAh g$^{-1}$ with a Coulombic efficiency of 99%. This represents approximately 40% of the maximum theoretical value of Si, long lifetime, and a specific capacity that is four times greater than the practical capacity achievable for commercial graphite electrodes, i.e., 372 mAh g$^{-1}$. To our knowledge, this capacity retention is superior to all previous reported results obtained for Si electrodes, which emphasizes the advantages of the nanostructured nature of Si combined with the carpet of VACNTs as efficient nanostructured current collector.[27,43,44]

## Rate capability retention

In order to evaluate the rate capability of the SiNPs@VACNTs electrode, galvanostatic measurement tests were performed at different C rates from C/20 to 10C. Figure 9b presents the specific capacities of the 30 μm VACNTs with 5 minutes Si deposition electrode obtained as a function of the lithiation / delithiation rate. The average Li de-alloying capacities at current rates of C/20, C/10, C/5, 1C, 5C, and 10C were 2500, 1750, 1400, 850, 250 and 100 mAh g$^{-1}$, respectively, and demonstrates an outstanding rate capability. Remarkably, a return to C/20 current rate either after 5C or 10C cycling rates, allowed recovering of an average 1450 mAh g$^{-1}$ capacity. The specific delithiation capacity decreases gradually with increasing the C rate, nevertheless, the capacities were found to be 2- up to 7-fold higher than the theoretical specific capacities of commercial graphite electrodes used in current LIBs. Furthermore, the possibility to recover the initial capacity value after lithiation / delithiation measurements at high C rates indicates a very good reversibility and no significant degradation of the SiNPs@VACNTs nanostructured anode when cycled at high rates.

The approach of SiNPs grafted to VACNTs enables thus the potential to better accommodate the associated huge volumetric change while maintaining the structural integrity of the electrode which has direct positive influence on the cycling stability. When the NPs start to form an alloy with lithium and undergo a large volume change, the anode can remain structurally intact because the highly conductive CNTs act as a flexible conductive wire mesh, allowing the electrochemically active nanoparticles to remain attached to the current collector of the anode and enhancing the electronic conductivity. Additionally, the fabrication path we propose herein avoids the extensive use of traditional solvents that are necessary during the regular electrode fabrication in LIBs industry, thus reducing the dead weight present in the state-of-the-art commercial anodes (15 to 20%).[45]

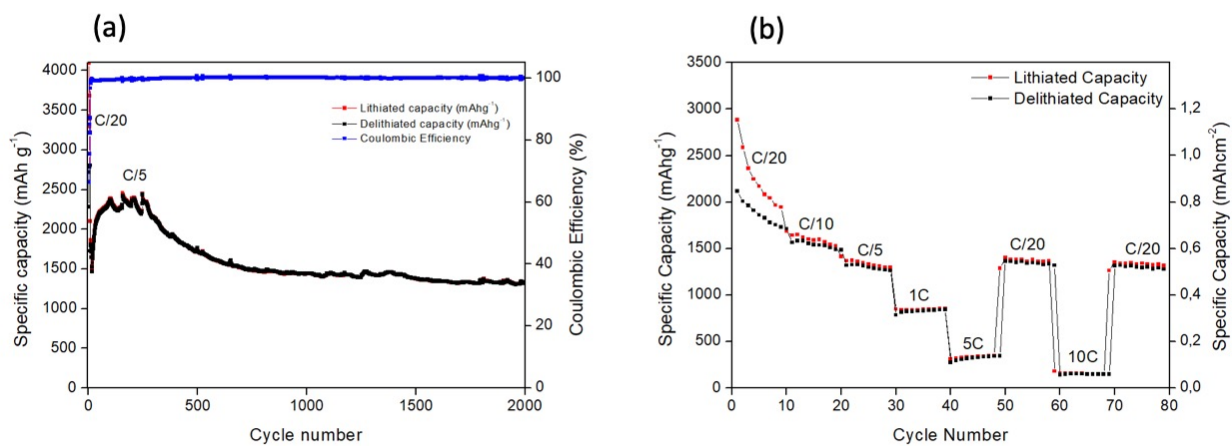

Figure 9. Lithiation / delithiation capacities and Coulombic efficiency of SiNPs@VACNTs electrode (30 µm of VACNTs and 5 min deposition of Si) during 2000 cycles at C/20 and C/5 rate (a) and rate capability retention of the electrode obtained at different lithiation/ delithiation rates (b).

## Conclusion

Hybrid hierarchical nanostructured SiNPs@VACNTs electrodes were directly synthesized on conventional Cu foil current collectors through simple and easily scalable manufacturing compatible processes, removing the use of polymer binder or carbon additives. The use of directly synthesized VACNTs on the Cu foil current collector ensured that every active SiNP is electrically connected onto the current collector through the VACNT carpet, thus providing an overall nanostructured electrode with a maximized electron and Li$^+$ ion transport. A first comparison on the particle size of the silicon revealed that the initial Coulombic efficiency values of such electrodes can be increased by increasing the SiNPs volume via the silicon deposition time. The general feature observed during the first galvanostatic cycle for the lighter loaded sample showed a larger irreversible capacity loss because of the larger amount of SEI being formed on the increased specific surface area of the Si nanostructures. However, by increasing the size of the SiNPs (by increased silicon deposition time), we were able to reduce the subsequent amount of the insulator SEI layer formed, which was translated into a higher Coulombic efficiency, but it negatively impacted the capacity stability of the electrode during the subsequent cycles. A second comparative approach, based on increasing the length of the VACNTs, was carried out for increasing silicon areal loading. The height of the VACNTs did not limit the electronic nor the Li$^+$ ions transport. Longer VACNTs allowed to achieve a higher gravimetric capacity than shorter ones at a given cycling current, as expected from the higher areal loading mass of Si. This study demonstrated similar electrochemical performances for both electrodes and showed a highly stable capacity over 40 cycles. Furthermore, we conducted thorough electrochemical testing of a lighter Si loaded SiNPs@VACNTs electrode with VACNT length of 30 µm. The reversible capacities obtained at C/5 rate after 2000 cycles retained 1330 mAh g$^{-1}$, which is more than four times higher than those of graphite anodes cycles at the same rate. This hierarchical hybrid nanostructure exhibited capacities to be 2- up to 7-fold higher than the theoretical specific capacities of commercial graphite electrodes used in current LIBs. Furthermore, this electrode sustained very high C rates without significant degradation of the integrity of the SiNPs@VACNTs nanostructured anode. These excellent electrochemical performances will hopefully contribute to bringing electrodes with high Si content into the market, enabling batteries with improved energy density, cost and safety, all of which are essential for a society based on renewable energy sources.

**Experimental Section**

**Fabrication of the hybrid SiNPs@VACNTs electrode**

A two-step manufacture method was adopted for the fabrication of SiNPs@VACNTs electrode directly on a copper (Cu) foil substrate as macroscopic current collector (see Figure S1 in Supporting Information). Prior to the VACNTs growth, a 50 nm thick aluminum oxide ($Al_2O_3$) porous layer followed by a 5 nm layer of iron (Fe) as CNTs' catalyst layer were deposited on the conventional Cu foil substrate by molecular beam evaporation (MBE) under high vacuum ($10^{-9}$ mbar). The $Al_2O_3$ layer acts as a barrier layer between the Fe and the Cu substrate preventing metal alloying. Due to the porous structure and rough surface of this layer, it uniformly pins the Fe catalyst in its surface-defects-trapping-sites, facilitating the subsequent growth of densely packed VACNTs. The VACNTs growth was carried out by a dHF-CVD technique in a horizontal tube furnace. [46] For each synthesis run, and before introducing $H_2$ and the precursor $CH_4$ gas, the system was heated to the desired growth temperature of 600 °C. Once set temperature was reached, to de-wet the catalyst layer into small Fe NPs, we carried a 5 minutes pretreatment step under activated $H_2$ introduced into the reactor at a flow rate of 100 sccm for 50 mbar pressure and filament power of 180 W [47,48] During this step, the Fe surface diffusion is strongly activated under the effect of the highly activated hydrogen that create/maintain trap sites on the $Al_2O_3$ layer that uniformly holds the Fe atoms on the surface, resulting in the formation of a homogeneously dispersed catalytic nanoparticle layer with small diameters and very high density.[49] Subsequently, the layer of NPs is exposed for 30 to 60 minutes to a mixture of $H_2/CH_4$ for the synthesis of densely packed VACNTs. The precursor gas mixture was added into the reactor at a pressure of 50 mbar (flow rate of $H_2$=50 sccm, $H_2$ filament power = 180 W, flow rate of $CH_4$=50 sccm, $CH_4$ filament power = 205 W). The contribution of $H_2$ during the CNTs growth is essential as it acts as an etching agent, thus limiting parasitic amorphous carbon deposition.[20,50,51] After the CNT growth, SiNPs were deposited on the VACNTs' walls via a CVD process at 540 °C using a mixture of 30 sccm of $H_2$ and 10 sccm of silane ($SiH_4$). Different deposition times, between 5 and 12 minutes, have been explored. The resulting loading mass of the Si active material was accurately determined by weighing the mass of the VACNT substrate before and after Si deposition using a quartz crystal microbalance. One can note that all the preparation sequences rely on well known, reliable and easily scalable and integrable processes inherited from microelectronics processing technology.

**Morphological, Chemical and Structural Characterization**

For morphology analyses of the hybrid nanostructured anodes, scanning electron microscopy (SEM) images were obtained using a HITACHI S 4800 FEG operating at 10 kV. Transmission electron microscopy (TEM) analyses were carried out in the conventional TEM mode using a TITAN G2 electron microscope operating at 300 kV. To confirm the composition of SiNPs, High-angle annular dark-field scanning transmission electron microscopy (HAADF-STEM) images and Energy-dispersive X-ray spectroscopy (EDS) were performed on a Titan Themis transmission electron microscope operating at 200 KV and equipped with a Cs aberration probe corrector and a Super X detector.

**Electrochemical Characterization**

For electrochemical tests, the hybrid nanostructured electrodes were directly assembled into CR2032 coin cells. The half-cell assembly was prepared in an argon-filled glove box using metallic lithium foil as counter electrode, a combination of glass fiber (Whatman, GF/C) and propylene (Celgard, 2400) as separators soaked in the electrolyte solution, and SiNPs@VACNTs as working electrode. Homemade electrolyte was prepared from 1M Lithium hexafluorophosphate ($LiPF_6$) in mixture (1:1 by volume) of ethylene carbonate (EC) and dimethyl carbonate (DMC), plus an additive of fluoroethylene carbonate (FEC 5%). All half-cells were kept idling for at least 6 hours at open circuit voltage before testing. Galvanostatic measurements were performed using a VMP3 (Bio-Logic) at room temperature. All cells were cycled in the potential window from 0.02 to 2 V vs $Li^+/Li$.

ACKNOWLEDGMENTS

The authors acknowledge financial support from Chaire EDF "Energies Durable", the Renault-ParisTech Sustainable Mobility Institute. The French state managed by the National Research Agency under the ANR 2018-CE09-0033 – "Harverstore" and the Investments for the Future program under the references ANR-10-EQPX-50 pole "Nano-Max & &NanoTEM". The authors would also like to acknowledge the Centre Interdisciplinaire de Microscopie électronique de l'X (CIMEX). The authors acknowledge the support of the X-ray crystallography facility, DIFFRAX, in Ecole Polytechnique, Institut Polytechnique de Paris. The authors thank Sandrine Tusseau-Nenez (PMC, Ecole Polytechnique, Institut Polytechnique de Paris) for her help for XRD data collection and their analyses. This work is part of the NanoMaDe-3E Initiative.

**Keywords**

**Table of contents:**

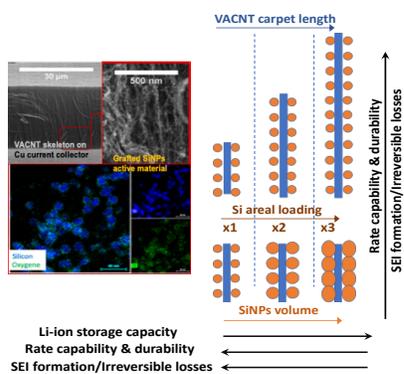

The exploitation of the combined merits of SiNPs and VACNTs. A hybrid nanostructured anode based on SiNPs anchored on VACNTs with defined spacing to accommodate volumetric changes of SiNPs. The cycling stability and the SEI formation of SiNPs@VACNTs is studied by tuning the Si areal loading either through the modulation of the SiNPs volume by changing the Si deposition time at a fixed VACNTs carpet length or through the variation of VACNT length at a fixed SiNPs volume.

**Supporting Information**

**The fabrication steps of the SiNPs@VACNTs electrode.**

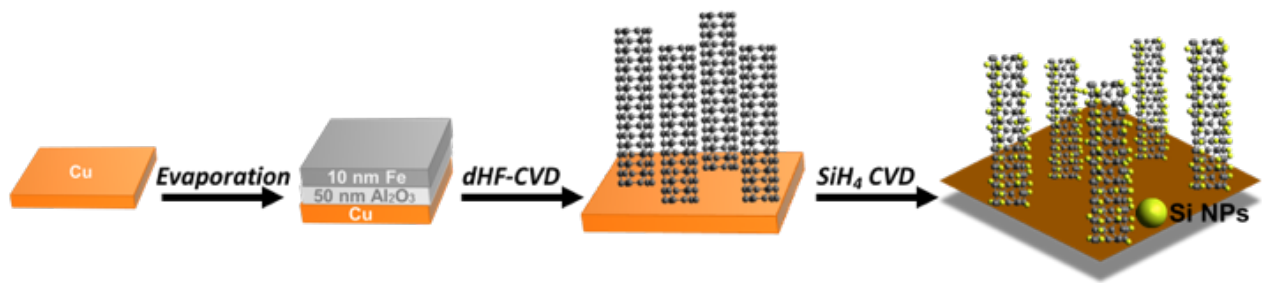

Figure S1. Schematic illustration of the fabrication of the SiNPs@VACNTs electrode on a copper foil.

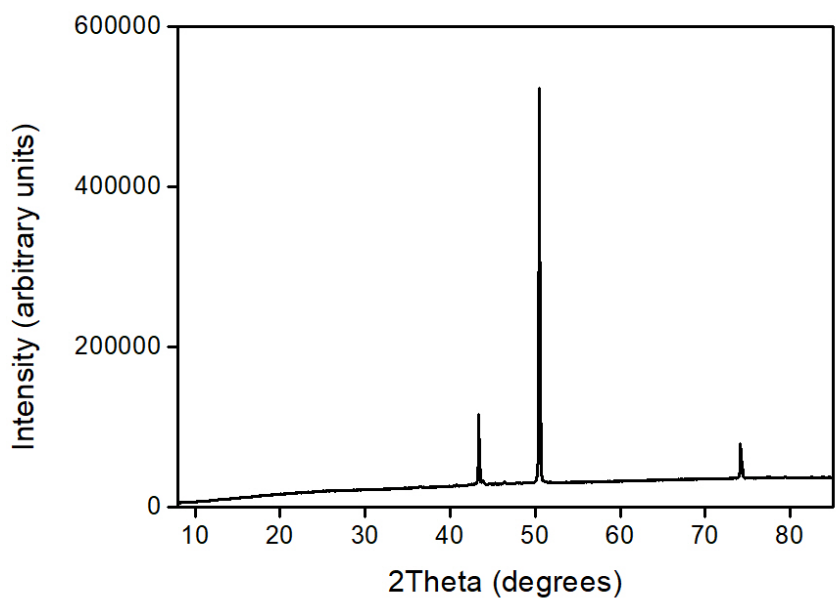

Figure S2. XRD spectrum of SiNPs@VACNTs on Cu foil.

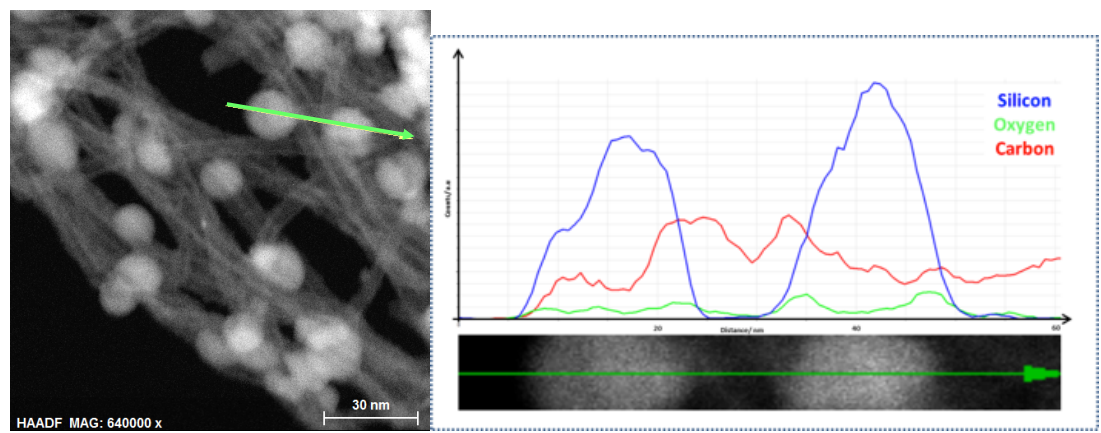

Figure S3. STEM-HAADF image of SiNPs@VACNTs and the corresponding STEM-EDS line scan spectra acquired along the region indicated by green arrow in the STEM-HAADF image.

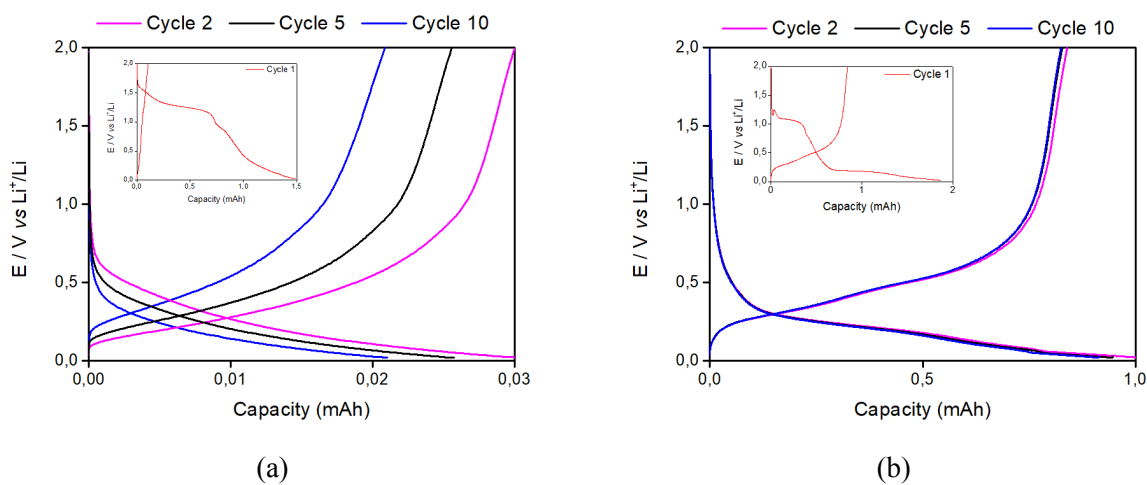

Figure S4. Galvanostatic plots for the 1st at C/20, and 2nd, 5th and 10th cycles at 1C of pristine VACNTs (a) and SiNPs@VACNTs (b) electrodes.

| Description | Mass loading (mg cm²) | Reversible capacity (mAh g⁻¹) based on the electrode | Areal capacity (mAh cm⁻²) | Reference |
|---|---|---|---|---|
| SiNPs@VACNTs | 0.2 – 2.5 | 1330 (after 2000 cycles) | 0.2 – 8.5 | This work |
| Si@CNT/C-microscroll | 0.8 | 2710 (after first few cycles) | 5.58 | Energy Environ. Sci. 13 (2020) 848-858 |
| VA-CNTs/Si | 0.17 | 2980 (after first few cycles) | 0.3 | Adv. Mater. 24 (2012) 2592–2597 |
| Si/CNT | 0.01 (total weight) | 2552 (after first few cycles) | - | ACS Nano. 4 (2010) 2233–2241 |
| Si-CNTs nanocomposite | 0.2 | 3112 (after first few cycles) | - | Nanoscale. 7 (2015) 3504–3510. |

| SiNWs electrode | 0.25 | 2900 (after first few cycles) | - | Electrochim. Acta. 157 (2015) 218–224 |
| Cu foam supported Si-based composite | 10 | 1200 (applied limitation on the capacity) | 10 | Adv. Energy Mater. 4 (2014) 1301718 |

Table S1. A summary of the performances of different Si-based anodes in comparison with the results in this work.

**Theoretical calculation of the areal capacity as a function of the length of the VACNTs and Si diameter**

In the following, we assume that the Si deposited on the walls of the CNTs form a continuous layer and that the total diameter of the cylinder composed of silicon covering the CNT after expansion (upon lithiation) is $D_{Total} = 50$ nm. We assume that the distance between two cylinders is d = 5 nm. Then, the total number of CNTs per cm$^2$ is:

$$\left(\frac{10^7}{D_{Total} + d}\right)^2 = 3.31 \times 10^{10}$$

Assume that the external diameter of the CNT is D = 8 nm and the height is H, the volume of each CNT is:

$$V_{CNT} = \pi \times \left(\frac{D}{2}\right)^2 \times H$$

If we assume that the diameter of the cylinder occupied by Si (before expansion) is $D_{Si}$, then the volume of the Si tube surrounding the CNT wall is:

$$V_{Si} = \left[\pi \times \left(\frac{D_{Si}}{2}\right)^2 \times H\right] - V_{CNT}$$

Therefore, the mass of Si in g per tube is:

$$m_{Si} = \frac{V_{Si} \times 2.3}{10^{21}}$$

where 2.3 is the density of silicon in g cm$^{-3}$ and $10^{21}$ is a unit conversion factor.

The total mass of silicon in g per cm$^3$ is:

$$M_{Si} = m_{Si} \times 3.31 \times 10^{10}$$

Therefore, the areal capacity (mAh cm$^2$) can be calculated from the theoretical capacity of silicon, i.e., 3580Ah/g according to

$$\text{Areal capaciy [mAh cm}^2] = M_{Si}[\text{mg cm}^2] \times 3.58[\text{mAh g}^{-1}]$$

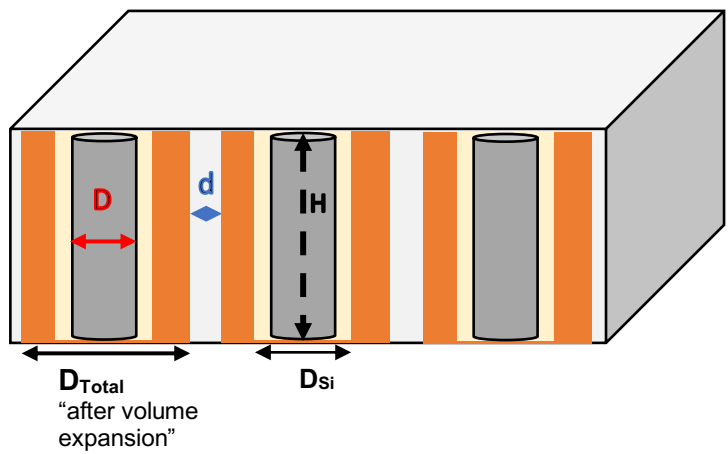